\def \SAIT #1 #2 {{\em Mem.\ Soc.\ Astron.\ It.\/} {\bf #1}, #2}
\def \MESS #1 #2 {{\em The Messenger\/} {\bf #1}, #2}
\def \ASTRNACH #1 #2 {{\em Astron. Nach.\/} {\bf #1}, #2}
\def \AAP #1 #2 {{\em Astron. Astrophys.\/} {\bf #1}, #2}
\def \AAL #1 #2 {{\em Astron. Astrophys. Lett.\/} {\bf #1}, L#2}
\def \AAR #1 #2 {{\em Astron. Astrophys. Rev.\/} {\bf #1}, #2}
\def \AAS #1 #2 {{\em Astron. Astrophys. Suppl. Ser.\/} {\bf #1}, #2}
\def \AJ #1 #2 {{\em Astron. J.\/} {\bf #1}, #2}
\def \ANNREV #1 #2 {{\em Ann. Rev. Astron. Astrophys.\/} {\bf #1}, #2}
\def \APJ #1 #2 {{\em Astrophys. J.\/} {\bf #1}, #2}
\def \APJL #1 #2 {{\em Astrophys. J. Lett.\/} {\bf #1}, L#2}
\def \APJS #1 #2 {{\em Astrophys. J. Suppl.\/} {\bf #1}, #2}
\def \APSS #1 #2 {{\em Astrophys. Space Sci.\/} {\bf #1}, #2}
\def \ASR #1 #2 {{\em Adv. Space Res.\/} {\bf #1}, #2}
\def \BAIC #1 #2 {{\em Bull. Astron. Inst. Czechosl.\/} {\bf #1}, #2}
\def \JSQRT #1 #2 {{\em J. Quant. Spectrosc. Radiat. Transfer\/} {\bf #1}, #2}
\def \MN #1 #2 {{\em Mon. Not. R. Astr. Soc.\/} {\bf #1}, #2}
\def \MEM #1 #2 {{\em Mem. R. Astr. Soc.\/} {\bf #1}, #2}
\def \PLR #1 #2 {{\em Phys. Lett. Rev.\/} {\bf #1}, #2}
\def \PASJ #1 #2 {{\em Publ. Astron. Soc. Japan\/} {\bf #1}, #2}
\def \PASP #1 #2 {{\em Publ. Astr. Soc. Pacific\/} {\bf #1}, #2}
\def \NAT #1 #2 {{\em Nature\/} {\bf #1}, #2}

\documentstyle{memsait}
\input epsf.sty
\begin{opening}
\title{UPPER LIMITS ON RADIO EMISSION FROM THE YOUNG X-RAY PULSARS 
IN THE SUPERNOVA REMNANTS G11.2$-$0.3 AND N157B}
\author{FRONEFIELD CRAWFORD$^1$, VICTORIA M. KASPI$^1$, 
RICHARD N. MANCHESTER$^2$, FERNANDO CAMILO$^3$, ANDREW G. LYNE$^3$, 
AND NICHI D'AMICO$^4$}
\institute{$^1$Department of Physics and Center for Space Research,
Massachusetts Institute of Technology, Cambridge, MA, USA\\
$^2$Australia Telescope National Facility, Epping, NSW, Australia\\
$^3$University of Manchester, Jodrell Bank, Macclesfield, Cheshire, UK\\
$^4$Osservatorio Astronomico di Bologna, Bologna, Italy}
\date{} % DO NOT INSERT ANY DATE HERE !!!
\end{opening}

\begin{document}

%\oddpagefooter{\sf Mem. S.A.It., Vol. ??, ??}{}{\thepage}
%\evenpagefooter{\thepage}{}{\sf Mem. S.A.It., Vol. ??, ??}
\oddpagefooter{}{}{} % LEAVE AS IT IS !
\evenpagefooter{}{}{} % LEAVE AS IT IS !
\ 
\bigskip

\begin{abstract}
Using the Parkes radio telescope, we have searched for pulsed radio
emission from the recently discovered X-ray pulsars AX J1811.5$-$1926
and PSR J0537$-$6910 in the supernova remnants G11.2$-$0.3 and N157B,
respectively.  We detected no significant pulsed radio emission from
these pulsars and have set an upper limit of 0.07 mJy on the 1374 MHz
flux from AX J1811.5$-$1926 and upper limits of 0.18 mJy and 0.06 mJy
for the flux from PSR J0537$-$6910 at 660 MHz and 1374 MHz,
respectively.  Assuming a power law radio spectral index of 2, these
flux limits correspond to luminosity limits at 400 MHz of 20 mJy
kpc$^{2}$ for AX J1811.5$-$1926 and 1100 mJy kpc$^{2}$ for PSR
J0537$-$6910.  Our luminosity limit for AX J1811.5$-$1926 is lower
than the observed luminosities of other young radio pulsars. The upper
limit on the luminosity for PSR J0537$-$6910 in N157B is not
significantly constraining due to the large distance to the pulsar. We
have also searched for giant radio pulses from both pulsars and have
found no convincing candidates.
\end{abstract}

\section{Introduction}
Recently, two rotation-powered X-ray pulsars were discovered in the
young supernova remnants G11.2$-$0.3 and N157B. The pulsar
AX J1811.5$-$1926 in the Galactic remnant G11.2$-$0.3, discovered in
archival {\it ASCA} data, has a 65 ms period (Torii et al. 1997). PSR
J0537$-$6910, discovered in N157B in the Large Magellanic Cloud (LMC),
was found in {\it XTE} data and has a period of 16 ms, making it the
fastest known pulsar associated with a remnant (Marshall et al. 1998).
The discovery of these two pulsars brings the total number of known
young Crab-like pulsars in plerionic nebulae to five and doubles the
number of such systems known in the LMC.

The discovery of pulsed radio emission from these systems would be
important for several reasons. A determination of the dispersion
measure (DM) for these pulsars would constrain the electron
distribution toward the LMC as well as the Taylor \& Cordes (1993)
Galactic DM-distance model. In addition, determining the radio
luminosity distribution for young pulsars is important for population
synthesis models.  Furthermore, ground-based radio telescope
facilities are often more practical for long-term timing of pulsars
than are X-ray instruments.  For young pulsars with high spin-down
rates, long-term timing can determine accurate braking indexes. In the
several cases where young pulsars show infrequent but sudden spin-ups
known as glitches, radio timing can also give information on the
interior physics of the neutron star (e.g. Shemar
\& Lyne 1996). In addition, measuring a time-of-arrival phase offset
between the radio and X-ray pulses helps constrain the pulsar magnetic field
geometry and radiation emission mechanism (e.g. Romani \& Yadigaroglu
1995).

\section{Observations and Analysis} 
We have undertaken a search for pulsed radio emission from these two
pulsars using the 64-m radio telescope in Parkes, NSW, Australia. We
conducted the search at two center frequencies, 660 MHz and 1374
MHz. For the 1374 MHz observations, we used the center beam of the
newly installed 20-cm multibeam receiver with a filter-bank back end
designed and built at Jodrell Bank. The search setup used was similar
to that of the current Parkes multibeam pulsar survey, described
elsewhere (Camilo et al. 1997). AX J1811.5$-$1926 was observed for a
single 4.5 hr integration at 1374 MHz while PSR J0537$-$6910 was
observed twice at 660 MHz in separate 4 hr integrations and once at
1374 MHz for 6 hr. Data were recorded on magnetic tape for processing
offline.  As part of this same project, we searched for and found the
pulsar PSR J1617$-$5055 at 1374 MHz near the supernova remnant
RCW 103 (Kaspi et al. 1998) which was also first discovered as an
X-ray pulsar from archival {\it ASCA} data (Torii et al. 1998).

Since the predicted topocentric period for AX J1811.5$-$1926 was
uncertain due to the unknown period derivative of the pulsar, we
performed a standard spectral analysis, after first dedispersing the
data at a variety of DMs from 0 to 1477 pc cm$^{-3}$. After each
dedispersion, the frequency channels were summed and a power spectrum
computed from the resulting time series. This power spectrum was then
searched for significant spikes. Each candidate frequency was then
further investigated by dedispersing and folding the original data at
the candidate DM and period. A search was then performed for a small
range in DM and period around the nominal DM and period and the
highest resulting signal-to-noise profile was computed.

Since the ephemeris for PSR J0537$-$6910 has a well-constrained period
derivative (Marshall et al. 1998), we folded the data at the predicted
topocentric period at a variety of DMs spanning 0 to 300 pc cm$^{-3}$,
which includes the expected DM range of 50 to 200 pc cm$^{-3}$ for LMC
pulsars, based on previous pulsar discoveries (McConnell et al. 1991,
Manchester et al. 1993).  In each DM trial, the data were searched in
a period range spanning the nominal folding period. For the 660 MHz
data, this period range was $\pm$150 ns, and for the 1374 MHz data the
range was $\pm$95 ns, corresponding to $\pm$36$\sigma$ and
$\pm$23$\sigma$ from the expected period, respectively.

We have also searched the dedispersed time series for Crab-like giant
radio pulses at the expected range of DMs. In this search, each
dedispersed time series was rebinned at a variety of widths and
searched for samples greatly exceeding the noise.  Potential giant
pulse candidates were investigated by examining the non-dispersed time
series for a strong signal in the same part of the time series in
order to rule out signals from terrestrial interference. For pulsar AX
J1811.5$-$1926, we searched a DM range from 40 to 670 pc cm$^{-3}$ and
found no convincing giant pulse candidates. For PSR J0537$-$6910, we
searched the two 660 MHz data sets and the 1374 MHz data set over the
expected LMC DM range of 50 to 200 pc cm$^{-3}$. We found no
convincing giant pulse candidates in any of the three N157B data sets.

\section{Results and Discussion}
We did not detect radio pulsations from either of the two X-ray
pulsars. By assuming distances to the remnants and using the Taylor \&
Cordes (1993) DM-distance model to estimate temporal smearing of
pulses from dispersion, we set upper limits for pulsed emission of
0.07 mJy and 0.06 mJy at 1374 MHz for AX J1811.5$-$1926 and PSR
J0537$-$6910, respectively. We have also set an upper limit of 0.18
mJy for the 660 MHz flux for PSR J0537$-$6910. By assuming a radio
power law index of $\alpha$ = 2 and distances of $d = 5$ kpc for
G11.2$-$0.3 (Green et al. 1988) and $d = 47$ kpc for N157B (Gould
1995), we obtain 400 MHz luminosity ($L_{400}$) upper limits for the
two pulsars, for comparison with other sources. For AX J1811.5$-$1926,
$L_{400}<20$~mJy kpc$^{2}$. For PSR J0537$-$6910, the luminosity upper
limit is $L_{400}<1100$ mJy kpc$^{2}$ using the more constraining 660
MHz flux limit. These results are outlined in Table 1.

\vspace{1cm} %TO ALLOW SUFFICIENT SPACE BETWEEN THE TEXT AND THE FIGURES
\centerline{\bf Table 1 - Summary of Radio Observations}
 
\begin{table}[h]
\hspace{1.8cm} %if you want to center your table act on this argument
\begin{tabular}{|l|c|c|}
\hline 
%\hline
Remnant Name           & G11.2$-$0.3        & N157B              \\
Pulsar Name            & AX J1811.5$-$1926  & PSR J0537$-$6910   \\
Period                 & 65 ms              & 16 ms              \\
$S_{660}$ upper limit  & --                 & 0.18 mJy           \\
$S_{1374}$ upper limit & 0.07 mJy           & 0.06 mJy           \\
Assumed distance       & 5 kpc              & 47 kpc             \\
$L_{400}$ limit (for $\alpha$ = 2) & 20 mJy kpc$^{2}$   & 1100 mJy kpc$^{2}$ \\
\hline 
\end{tabular}
%%%%%%%%%%%%%%%%%%%%%%%%%%%%%%%%%%%%%%%%%%%%%%%%%%%%%%%%%%%%%%%%%%%%%%%%%
\end{table}
%%%%%%%%%%%%%%%%%%%%%%%%%%%%%%%%%%%%%%%%%%%%%%%%%%%%%%%%%%%%%%%%%%%%%%%%%%

Of the five known radio pulsars having characteristic age $\tau_{c}
\equiv P/2\dot{P} < 10$ kyr (the Crab pulsar, PSR B1509$-$58, 
PSR B0540$-$69, PSR B1610$-$50, and PSR J1617$-$5055), PSR B1509$-$58
has the lowest pulsed radio luminosity, $L_{400} \simeq 50$~mJy
kpc$^{2}$.  The largest luminosities belong to the Crab pulsar and PSR
B0540$-$69, which have $L_{400} = 2600$~mJy kpc$^{2}$ and $1800$~mJy
kpc$^{2}$ respectively. Of course, all luminosity estimates are
distance-dependent, hence subject to some uncertainty.

Our upper limit of $L_{400} < 20$ mJy kpc$^{2}$ for AX J1811.5$-$1926
is significantly lower than the luminosities of the Crab pulsar and
PSR B0540$-$69 and is below the value for PSR B1509$-$58. This
suggests that AX J1811.5$-$1926 should have detectable radio
emission. From the Taylor \& Cordes (1993) DM-distance model, we
estimate that the expected DM for AX J1811.5$-$1926 is $\sim$350 pc
cm$^{-3}$, corresponding to about 3 ms dispersion smearing in each
frequency channel, and that the multipath scattering from plasma
inhomogeneities should be less than 1 ms. Thus, neither of these
effects should significantly affect the detectability of
the 65~ms periodicity.  Therefore, the
non-detection means that either the radio beam does not intersect our
line of sight or that the radio luminosities of very young pulsars can
be lower than what has been generally thought (e.g. Narayan \&
Ostriker 1990). If the latter is true, then deeper radio searches with
higher sensitivity may reveal a previously undiscovered population of
young faint pulsars.

Our luminosity limit is not as stringent for PSR J0537$-$6910 due to
its large distance. Also, the poorly known interstellar/intergalactic
medium in this direction could significantly scatter radio pulses
beyond detectability, particularly at 660~MHz.

\section{Conclusions}
We conclude from our search that the X-ray pulsar AX J1811.5$-$1926 in
G11.2$-$0.3 has pulsed radio emission that is certainly of
significantly lower flux than all other known pulsars of comparable
age, and is probably either underluminous relative to these sources or
is not beaming toward Earth. Our much less stringent limits on the
radio luminosity for PSR J0537$-$6910 in N157B are
unconstraining. Next generation telescopes such as the square
kilometer array would be ideal instruments to carry out deeper
searches toward these sources.

% References. 

\end{document}